\documentclass[a4paper,notitlepage,12pt]{article}

\begin{document}
\title{Periodic correlation structures in  bacterial and archaeal complete genomes}
\author{ Zuo-Bing Wu\footnotemark[1]\\
State Key Laboratory of Nonlinear Mechanics,\\
 Institute of Mechanics,\\
  Chinese Academy
of Sciences, Beijing 100190, China}
 \maketitle

\footnotetext[1]{Correspondence to. Tel.: 86-10-82543955. Email
address: wuzb@lnm.imech.ac.cn}

\newpage
\begin{abstract}
The periodic transference of nucleotide strings in bacterial and archaeal
complete genomes is investigated by using the metric representation and
the recurrence plot method. The generated periodic correlation structures
exhibit four kinds of fundamental transferring characteristics:
a single increasing period, several increasing periods, an increasing
quasi-period and almost noincreasing period. The mechanism of
the periodic transference is further analyzed by
determining all long periodic nucleotide strings in the bacterial
and archaeal complete genomes and is explained as follows: both the
repetition of basic periodic nucleotide strings and the
transference of non-periodic nucleotide strings would form the periodic
correlation structures with approximately the same increasing periods.

\end{abstract}

\newpage
\section{Introduction}
Since complete genomes of many organisms are available from
web-based databases, a full and systematic search of genome
structures, functions and dynamics becomes an essential part of the study
for both biologists and physicists. For the large amount of
genomes, developing quantitative methods to extract meaningful
information is a major challenge with respect to applications of statistical
mechanics and nonlinear dynamics to biological
systems\cite{Wille,Peyrard}. To understand the complete genomes,
some statistical and geometrical methods were
developed\cite{Jeffrey,Li,Peng,DGVFF,Hao,QC,KTP,RRP,SG,MAL,Kat,PO,GGS,AV,PK,Car}.
The studies of the complete genomes of many
organisms came up with the determinations of the nontrivial statistical
characteristics, such as the long-range correlations, the short-range
correlations and the fractal features or genomic signatures. In
particular, it was found that the transposable elements,
 as the mobile DNA sequences, have the ability to move from one place to another and
make many replicas within the genome via the
transposition\cite{BEN,FJW,Kazazian}. Their origin,
evolution, and tremendous effects on the genome structure and the gene
function are issues of fundamental importance in biology\cite{BQ,LS,Sha}.

In general, the symbolic dynamics and the recurrence plots are basic
methods of nonlinear dynamics for analyzing complex
systems\cite{HaoS,Kurth}. Although the conventional methods have made
great strides in understanding genetic patterns, they are required
to analyze the so-called junk DNA with complex functions governing
mutations\cite{Guastello,Spinelli}. Recently, a one-to-one metric
representation of a genome borrowed from the symbolic dynamics was
proposed to form a fractal pattern in a plane\cite{Wu1,Wu2}. By
using the metric representation method, the recurrence plot
technique of the genome was established to analyze the correlation
structures of nucleotide strings\cite{Wu3,Wu4}. The transference
of nucleotide strings appears at many positions of a complete
genome and makes a regular and irregular correlation structures,
but the periodic correlation structures in the complete genome
are the most interesting in view of the dynamics. In this paper,
using the metric representation and the recurrence plot method, we
identify periodic correlation structures in bacterial and
archaeal complete genomes and analyze the mechanism of the
periodic correlation structures. Since the nucleotide strings include
transposable elements, the mechanism is conducible to understanding the
genome structures in terms of nucleotide strings transferring in the
genomes and exploring relations between transference of nucleotide strings
and the transposable elements.

\section{Correlation structures in periodic and random sequences}
In what follows, we give a brief presentation of the metric
representation and the recurrence plot method, which are detailed in
\cite{Wu1,Wu2,Wu3,Wu4}.
For a given symbolic sequence $s_1 s_2 \cdots s_i
\cdots s_N$ ($s_i \in \{A,C,G,T\}$), a metric representation
for its subsequences $\Sigma_k=s_1 s_2 \cdots s_k$ ($1 \le k \le N$)
is defined as
 \begin{equation}
 \begin{array}{l}
 \alpha_k  = 2\sum_{j=1}^k \mu_{k-j+1} 3^{-j} +3^{-k}= 2\sum_{i=1}^k \mu_i 3^{-(k-i+1)}
 +3^{-k},\\
 \beta_k = 2\sum_{j=1}^k \nu_{k-j+1} 3^{-j} +3^{-k}= 2\sum_{i=1}^k \nu_i 3^{-(k-i+1)} +3^{-k},
 \label{eq1}
 \end{array}
 \end{equation}
 where $\mu_i$ is 0 if $s_i \in \{A,C\}$ or 1 if $s_i \in \{G,T\}$
 and $\nu_i$ is 0 if $s_i \in \{A,T\}$ or 1 if $s_i \in \{C,G\}$.
 It maps the one-dimensional symbolic sequence to
 the two-dimensional plane ($\alpha, \beta$).
The subsequences with the
same ending $l$-nucleotide string are labeled with
$\Sigma^{l}$. They correspond to points in the zone encoded by the
$l$-nucleotide string.
With two subsequences $\Sigma_i \in
\Sigma^{l}$ and $\Sigma_j$ ($j \geq l$), we calculate
\begin{equation}
\Theta(\epsilon_{l}-|\Sigma_i-\Sigma_j|)= \Theta(\epsilon_{l}-
\sqrt{(\alpha_i-\alpha_j)^2+(\beta_i-\beta_j)^2}), \label{eq3}
\end{equation}
where $\epsilon_{l}=1/3^{l}$ and $\Theta$ is the Heaviside
function [$\Theta(x)=1$, if $x > 0$; $\Theta(x)=0$, if $x \leq 0$].
 When
$\Theta(\epsilon_{l}-|\Sigma_i-\Sigma_j|)=1$, i.e., $\Sigma_j \in
\Sigma^{l}$, a point $(i,j)$ is plotted on a plane.
Repeating the above process for $i \in [l,N]$ and $j \in [l,N]$,
 we obtain a recurrence plot of the symbolic sequence.
To present the correlation structure
 in the recurrence plot plane, we define a correlation intensity
 at a given correlation distance $d$ as
\begin{equation}
\Xi(d) = \sum_{i=1}^{N-d}
\Theta(\epsilon_{l}-|\Sigma_i-\Sigma_{i+d}|), \label{eq4}
\end{equation}
which displays the transference of $l$-nucleotide strings
in the symbolic sequence.
On the recurrent plot plane, since $\Sigma_i$ and $\Sigma_j$ $\in \Sigma^{l}$,
the transferring element has a length $l$
at least. We calculate the maximal value of $x$ to satisfy
\begin{equation}
\Theta(\epsilon_{l}-|\Sigma_{i+x}-\Sigma_{j+x}|)=1, \ \ \
x=0,1,2,\cdots x_{max},\label{eq5}
\end{equation}
i.e., $\Sigma_{i+x}$ and $\Sigma_{j+x} \in \Sigma^l$. The
transferring element has a length $L=l+x_{max}$ and is placed at
the positions $(i-l+1, i+x_{max})$ and $(j-l+1, j+x_{max})$, which
implies the correction distance $d=j-i$.

To understand the transferring characteristics of a complex genome, we
investigate the correlation structures of simple periodic and random
sequences. By randomly combining the four letters A, C, G and T,
we firstly generate two random nucleotide sequences: one has the
length of 67 and another has the length of 5000. Then, a periodic
nucleotide sequence with the total length of 5000 is formed by
repeating the short nucleotide string. Using the metric
representation and the recurrence plot method, we may determine
the correlation intensities at different correlation distances with
$l=8$ for the periodic and random sequences in Fig.~1. It is
evident that there exist equidistant parallel lines with a
basic correlation distance in Fig.~1(a), to form the periodic
correlation structure for the periodic sequence. The basic
correlation distance hereinafter called the basic periodic
length is determined as $d_b=67$. The correlation intensity
$\Xi(d)$ decreases linearly with the increase of the correlation
distance ($d_b$, $2d_b$, $\cdots$). However, in Fig.~1(b), the
correlation intensity $\Xi(d)$ is very small, so there are
almost no correlation structures for the random sequence.
Therefore, the periodic and random sequences exhibit two very
different transferring characteristics: with the periodic correlation
structure with a linearly decreasing intensity and without a clear
correlation structure.

\section{Periodic nucleotide strings in bacterial and archaeal complete genomes}

At the end of 1999, complete genomes including more of 20 bacteria
were in the Genbank\cite{Hao}.
By using the string composition and the metric representation method,
the suppressions of all short strings in 23 bacterial and archaeal
complete genomes were determined\cite{Hao,Wu2}.
In this section, using the metric representation and the recurrence
plot method, we determine all long periodic nucleotide strings ($
\geq 20$ bases) in the 23 bacterial and archaeal genomes.  For the 23
 genomes, only 13 have long periodic
nucleotide strings. All basic
strings and their lengths of the long periodic nucleotide strings
in the 13 bacterial and archaeal genomes are presented in Table I
in the order of decreasing suppressions of nucleotide
strings\cite{Wu2}. Several periods and different
basic strings can be seen depending on the genomes, but not necessarily
on the lengths of genomes. The genomes of Helicobacter
pylori 26695 ($hpyl$), Helicobacter pylori J99 ($hpyl99$),
Haemophilus influenzae Rd KW20 ($hinf$), Mycobacterium tuberculosis H37Rv
($mtub$), Synechocystis sp. PCC6803 ($synecho$) have more periods ($
\geq 6$) and basic strings ($ \geq 9$) than others, which have only fewer
periods ($ \leq 3$) and basic strings  ($ \leq 4$). In each
period, the number of the basic strings generally depends on the
length of the period. The longer/shorter period the basic strings
have, the smaller/greater their number will be. In the next section, we will
investigate the periodic transference of nucleotide strings in the bacterial
and archaeal complete genoms and analyze the effects of periodic
nucleotide strings on the correlation structures.

\section{Periodic correlation structures in bacterial and archaeal complete genomes}

The periodic correlation structures of a complete genome contain
several basic periodic and/or quasi-periodic lengths, which are
determined by using the metric representation and the recurrence plot
method as follows. From the relationship between the correlation intensity
and the correlation distance obtained by using Eq. (3), the basic
periodic lengths and their integer multiples with strong
correlation intensities can be calculated. Moreover, in the
transference of nucleotide strings obtained by using Eq. (4),
the correlation distance with basic periodic lengths and their integer
multiples can also be found. By using both methods, the basic
periodic lengths of the periodic correlation structures are
determined, as shown in Table II, where the 23 complete
genomes with official genbank accession numbers
are arranged in the order of decreasing suppressions of
nucleotide strings\cite{Wu2}. When the periodic correlation
structures have only a few peaks of the correlation intensity within
the correlation distance, the basic periodic lengths are put in parentheses.
To see the characteristics of the periodic
correlation structures, we also present all basic string lengths
in long periodic nucleotide strings ($ \geq 20$ bases) in Table
II. When a periodic correlation structure is identified based on a
long periodic nucleotide string, the transference of nucleotide
strings composed of the basic strings appears at some positions
where the correlation distance is integer multiples of the period
and monotonically increases. At the same time, the lengths of
transferred nucleotide strings monotonically decrease. There
exists a "cascade" arrangement of nucleotide strings related to
the basic periodic length. However, when a periodic correlation
structure is identified based on non-periodic nucleotide strings,
the transference of nucleotide strings appears at several
positions where the correlation distance is almost integer multiples of
the basic periodic length. There are no "cascade"
arrangements of nucleotide strings related to the basic periodic
length. According to the characteristics of the periodic correlation
structures, the results can be summarized as follows:

(1)The correlation distance contains a single increasing period. The most of the
complete genomes with a single increasing period
have a basic periodic length of 67. They include Methanococcus
jannaschii DSM 2661 ({\it mjan}), Methanobacterium thermoautotrophicum str. delta H
({\it mthe}), Pyrococcus horikoshii OT3 ({\it pyro}), Archaeoglobus
fulgidus DSM 4304 ({\it aful}), Pyrococcus abyssi ({\it pabyssi}) and
Thermotoga maritima MSB8 ({\it tmar}) genomes. Consider the {\it
mjan} genome as an example. Fig. 2 displays the
correlation intensity at different correlation distances with
$l=15$ for the {\it mjan} genome. It is evident that there
exist some equidistant parallel lines with a basic periodic
length, to form a periodic correlation structure. The basic
periodic length is determined as $d_b=67$. Generally, if the
genome has a periodic nucleotide string with the basic string
length $p=d_b$, it would tend to form a periodic
correlation structure. In Table II, the {\it mjan} genome has the
correspondent basic string length $p=d_b$ for periodic nucleotide
strings. For example, the nucleotide string $\Sigma_1=at^2 \cdots
a^2t$ (237122-237620) with $L_{\Sigma_1}=499$ is formed by repeating
the basic string $at^2 \cdots t^2c$ with the length $p=d_b$, where
$L_{\Sigma_1}=7p+30$. In other words, the basic string is duplicated
to the positions with the correlation distances $p$, $2p$, $3p$,
$4p$, $5p$, $6p$ and $7p$. Despite possible contribution from
such periodic nucleotide strings, the periodic correlation structure is
mainly formed by the transference of non-periodic nucleotide
strings, which has approximately the same increasing period. For example, the
nucleotide string $\Sigma_2=a^2t^2a^4tcagac^2gt^3cg^2a^2tg^2a^3$ (447-476) with
$L_{\Sigma_2}=30$ is transferred to the places (514-543), (581-610),
(651-680), (718-747), (785-814), (855-884), (922-951), (994-1023),
(1064-1093), (1132-1161) and $\cdots$ with the correlation distances $d_b$, $2d_b$,
$3d_b+3$, $4d_b+3$, $5d_b+3$, $6d_b+6$, $7d_b+6$, $8d_b+11$, $9d_b+14$,
$10d_b+15$ and $\cdots$, respectively. Since the nucleotide string $\Sigma_2$
is neither periodic nor a part of a periodic nucleotide string,
its periodic transference is not a
repetition of basic periodic nucleotide strings.
Moreover, Fig.~2 shows that there also exists a cluster of
basic periodic lengths close to $d_b$. Their integer multiples
are distributed near the periodic correlation structure.
Table II shows that there also exists another basic string length $p=68$ for
periodic nucleotide strings, which is conducible to form the cluster
distribution near the periodic correlation structure. So both the
repetition of basic periodic nucleotide strings and the
transference of non-periodic nucleotide strings would form the periodic
correlation structure with approximately the same increasing period.

Besides the {\it mjan} genome, the other genomes ({\it mthe}, {\it
pyro}, {\it aful}, {\it pabyssi} and {\it tmar}) have no
periodic nucleotide strings with the basic string length $p=d_b$
to make contributions to the periodic correlation structure. So
the periodic correlation structure
is formed by the transference of non-periodic nucleotide strings.
Furthermore, the genomes of Mycoplasma genitalium G37 ({\it mgen}),
{\it hinf}, Mycoplasma pneumoniae M129 ({\it mpneu}), Treponema
pallidum subsp. pallidum str. nichols ({\it tpal}), Aeropyrum pernix K1 ({\it aero}), Rickettsia
prowazekii str. madrid E ({\it rpxx}) and Borrelia burgdorferi
B31 ({\it bbur}) have
basic periodic lengths $d_b=3$, 4, 12, 24, 65, 84 and 162,
respectively. In Table II, they correspond to periodic
nucleotide strings with the basic length $p=d_b$ except the {\it
aero} genome. So both the repetition of basic periodic nucleotide
strings and the transference of non-periodic nucleotide strings would form
the periodic correlation structure
 with approximately the same increasing period.

(2) The correlation distance contains several increasing periods.  The
Escherichia coli K-12 MG1655 ({\it ecoli}) genome has two basic periodic
lengths 100 and 113. The {\it hpyl99} genome has three basic
periodic lengths 8, 15 and 21. The {\it mtub} genome has three
basic periodic lengths 9, 15 and 57. Consider the {\it hpyl99}
genome as an example. Fig. 3 displays the correlation
intensity at different correlation distances with $l=15$ for the
{\it hpyl99} genome. It is evident that there exist some
equidistant parallel lines with basic periodic lengths, to form
periodic correlation structures. Three basic periodic lengths are
determined as $d_{b_1}=8$, $d_{b_2}=15$ and $d_{b_3}=21$. Although
there are some peaks of the correlation intensity in the
correlation distance as shown in Fig. 3, they do not form any periodic
correlation structures and are not accounted. Table II
also shows some periodic nucleotide strings with basic string
lengths $p_1=d_{b_1}$, $p_2=d_{b_2}$, $p_3=d_{b_3}$ and their integer
multiples, which contribute to the periodic correlation
structures. For example, the nucleotide string $\Sigma_1=ca^2
\cdots ca^2$ (1061079-1061153) with $L_{\Sigma_1}=75$ is formed by
repeating the basic string $ca^2c^2at^2$ with the length $p_1$,
where $L_{\Sigma_1}=9p_1+3$. The nucleotide string $\Sigma_2=a^3
\cdots ca^2$ (5153-5280) with $L_{\Sigma_2}=128$ is formed by
repeating the basic string $a^5ca^3ga^2t^3$ with the length $p_2$,
where $L_{\Sigma_2}=8p_2+8$. The nucleotide string $\Sigma_3=atc
\cdots tca$ (659300-659450) with $L_{\Sigma_3}=151$ is formed by
repeating the basic string $atcata^2t^2a^2c^3tca^3tc$ with the
length $p_3$, where $L_{\Sigma_3}=7p_3+4$. Although the
transference of non-periodic nucleotide strings might also
contribute to the periodic correlation structures, they are mainly
formed by repeating the basic periodic nucleotide strings.
For example, the non-periodic
nucleotide string $\Sigma_4=aga^4c^2ta^2cta^3ga^6c$ (59514-59537) with
$L_{\Sigma_4}=24$ is transferred to the places (59640-59663) and (59724-59747)
with the correlation distances $6d_{b_3}$ and $10d_{b_3}$, respectively.
So both
the repetition of basic periodic nucleotide strings and the
transference of non-periodic nucleotide strings would form the periodic
correlation structures with approximately the same increasing periods.

(3) The correlation distance has an increasing quasi-period.  The
Bacillus subtilis subsp. subtilis str. 168 ({\it bsub}) genome has a basic quasi-periodic
length of 5000.  Fig. 4 shows the correlation intensity at
different correlation distances with $l=15$ for the {\it bsub}
genome. It is evident that there exist some approximately
equidistant parallel lines at the positions $d=4996$, 10605, 15427
and 20468, to form a quasi-periodic correlation structure with a
basic quasi-periodic length $d_b \approx 5000$. Although a
stronger correlation intensity appears at the position $d=5856$,
it is far away from the quasi-periodic correlation structure and
is not accounted.
In Table II, there are no periodic
nucleotide strings with the length $p =d_b$ to make a contribution
to the quasi-periodic correlation structure.
For example, the non-periodic
nucleotide string $\Sigma_1=agc \cdots tac$ (167978-169382) with
$L_{\Sigma_1}=1405$ is transferred to the place (172974-174378)
with the correlation distance $4996$.
The non-periodic
nucleotide string $\Sigma_2=t^3 \cdots ct^2$ (161449-161666) with
$L_{\Sigma_2}=218$ is transferred to the places (167057-167274),  (172056-172273) and (946761-946798)
with the correlation distances $5608$, $10607$ and $785321$, respectively.
So the transference
of non-periodic nucleotide strings would form the quasi-periodic
correlation structure.

(4) The correlation distance contains a combination of several increasing
periods and an increasing quasi-period. Firstly, the {\it hpyl}
genome has two basic periodic lengths 7, 8 and a basic
quasi-periodic length of 114. Fig.~5(a) shows the correlation
intensity at different correlation distances with $l=15$ for the
{\it hpyl} genome, with a local region magnified. It
is evident that there exist some equidistant parallel
lines with basic periodic lengths, to form periodic correlation
structures in a short range of the correlation distance. The two
basic periodic lengths are determined as $d_{b_1}=7$ and $d_{b_2}
=8$. Moreover, in Fig.~5(a), there also exist some approximately
equidistant parallel lines at the positions $d=96$, 207, 324, 438,
552, 666 and 780, to form a quasi-periodic correlation structure
in a long range of the correlation distance. The quasi-periodic
correlation distance is described as $d \approx 96 + x d_{b_3}$,
where the basic quasi-periodic length $d_{b_3}$ is 114 and $x = 0,
1, 2, \cdots$. In Table II, there exist some periodic nucleotide
strings with basic string lengths $p_1=d_{b_1}$, $p_2=d_{b_2}$ and
their integer multiples, but no periodic
nucleotide strings with the basic string length $p_3=d_{b_3}$. For
example, the nucleotide string $\Sigma_1=tga \cdots t^2a$ (1-181)
with $L_{\Sigma_1}=181$ is formed by repeating the basic string
$tgat^2ag$ with the length $p_1$, where $L_{\Sigma_1}=25p_1+6$.
The nucleotide string $\Sigma_2=t^2g \cdots tga$ (444403-444490)
with $L_{\Sigma_2}=88$ is formed by repeating the basic string
$t^2gct^2ga$ with the length $p_2$, where $L_{\Sigma_2}=11p_2$.
Although the
transference of non-periodic nucleotide strings might also
contribute to the periodic correlation structures, they are mainly
formed by repeating the basic periodic nucleotide strings.
For example, the non-periodic
nucleotide string $\Sigma_3=atgtatg^2catg^3catgtatg$ (84905-84926) with
$L_{\Sigma_3}=22$ is transferred to the place (84929-94850)
with the correlation distance $3d_{b_2}$. Moreover, for the
quasi-periodic correlation structure, the non-periodic
nucleotide string $\Sigma_4=gtgagta^2t^3ctcgcat^2ct^6ctc$ (556196-556224) with
$L_{\Sigma_4}=29$ is transferred to the places (556634-556662),
(556748-556776), (556862-556906), (557300-557328), (557414-557442) and (557852-557880)
with the correlation distances $96+3d_{b_3}$, $96+4d_{b_3}$, $96+5d_{b_3}$, $78+9d_{b_3}$, $78+10d_{b_3}$ and $60+14d_{b_3}$, respectively.
So both the repetition of basic periodic nucleotide strings and the
transference of non-periodic nucleotide strings would form the periodic
correlation structures with approximately the same increasing periods in a
short correlation distance, but only the transference of
non-periodic nucleotide strings would form the quasi-periodic
correlation structure in a long correlation distance.

Secondly, the {\it synecho} genome has two basic periodic lengths
6, 888 and a basic quasi-periodic length of 296. Fig.~5(b) shows
the correlation intensity at different correlation distance with
$l=15$ for the {\it synecho} genome, with a local region magnified.
It is evident that there exist some
equidistant parallel lines with basic periodic lengths, to form
periodic correlation structures in short and long ranges of
correlation distances, respectively. Two basic periodic lengths are
determined as $d_{b_1}=6$ and $d_{b_2} = 888 $. Moreover, in
Fig.~5(b), there also exist some approximately equidistant
parallel lines at the positions $d_1 = 297 + x_1 d_{b_2} \approx
(1+3x_1)d_{b_3}$ and $d_2 = 591 + x_2 d_{b_2} \approx
(2+3x_2)d_{b_3}$, where the basic quasi-periodic length $d_{b_3}$
is 296 and $x_1, x_2=0, 1, 2, \cdots$. They form quasi-periodic
correlation structures in a long range of the correlation
distance. In Table II, there exist some periodic nucleotide
strings with integer multiples of $d_{b_1}$ and the basic string
length $p_2=d_{b_2}$, but no periodic nucleotide
strings with the basic string length $p_3=d_{b_3}$. For example,
the nucleotide string $\Sigma_1=gag \cdots tga$ (527703-527770)
with $L_{\Sigma_1}=68$ is formed by repeating the basic string
$gagc^2g^2a^2c^2tga^2c^3$ with the length $p_1=3d_{b_1}$, where
$L_{\Sigma_1}=3p_1+14$. The nucleotide string $\Sigma_2=cac \cdots
g^2t$ (2354010-2355833) with $L_{\Sigma_2}=1824$ is formed by
repeating the basic string $cac \cdots cat$ with the length $p_2$,
where $L_{\Sigma_2}=2p_2+48$. Moreover, the nucleotide string
$\Sigma_3=(ctga^2c^3gagc^2g^2a^2c)^2ctga$ (527395-527434) with $L_{\Sigma_3}=40$
is transferred to the places (527473-527512), (527491-527530),
(527509-527548), (527527-527566) and (527545-527584) with the
correlation distances $13d_{b_1}$, $16d_{b_1}$, $19d_{b_1}$,
$22d_{b_1}$, and $25d_{b_1}$, respectively.
 The non-periodic nucleotide string $\Sigma_4=cac \cdots tcg$
(2354010-2354300) with $L_{\Sigma_4}=291$ is transferred to the
places (2356674-2356964), (2357562-2357852) and (2358450-2358740)
with the correlation distances $3d_{b_2}$, $4d_{b_2}$ and
$5d_{b_2}$, respectively.
 Both the
repetition of basic periodic nucleotide strings and the
transference of non-periodic nucleotide strings would form the periodic
correlation structures with approximately the same increasing periods
in short and long correlation distances, but only the transference of
non-periodic nucleotide strings would form the quasi-periodic
correlation structures in a long correlation distance.

(5) The correlation distance contains almost no increasing
periods. The genomes of Aquifex aeolicus VF5 ({\it aquae}), Rhizobium
sp. NGR234 plasmid pNGR234a ({\it pNGR234}), Chlamydophila pneumoniae CWL029 ({\it cpneu}) and
Chlamydia trachomatis D/UW-3/CX ({\it ctra}) are among cases with such characteristics.
Consider the {\it aquae} genome as an example. Fig. 6 shows
the correlation intensity at different correlation
distances with $l=15$ for the {\it aquae} genome. It is
evident that there exist some equidistant parallel lines with a
basic periodic length, which is determined as
$d_b=67$. However, for the basic periodic length $d_b=67$, the
maximal correlation intensity $\Xi(d)$ is only 179 and the
correlation structure has only three peaks of the correlation
intensity at the positions $d_b$, $2d_b$ and $3d_b$. The weak
correlation intensity with a few peaks in the correlation
distance may not make any periodic correlation structures.
In Table II, there are also no periodic nucleotide
strings for the almost non-periodic correlation structure. So the
{\it aquae} genome almost has no periodic correlation
structures.

\section{Conclusion and discussions}
In summary, using the metric representation and the recurrence plot
method, we have observed periodic correlation structures in
bacterial and archaeal complete genomes. All basic periodic
lengths in the periodic correlation structures are determined. On
the basis of the periodic correlation structures, the bacterial
and archaeal complete genomes, as classified into five groups,
display four kinds of fundamental transferring
characteristics: a single increasing period, several increasing
periods, an increasing quasi-period and almost noincreasing
period. The mechanism of the periodic correlation structures
is further analyzed by determining all long
periodic nucleotide strings in the bacterial and archaeal complete
genomes and is explained as follows: both the repetition of basic
periodic nucleotide strings and the transference of non-periodic
nucleotide strings would form the periodic correlation structures with
approximately the same increasing periods.

In comparison with the complete genome of the Saccharomyces
cevevisiae yeast\cite{Wu4}, it is found that the bacterial,
archaeal and yeast complete genomes have the same four kinds of
fundamental transferring characteristics of nucleotide
strings. They choose preferably the basic periodic length $d_b
\approx 67$ or its double $d_b \approx 135$ in the periodic
correlation structures, even they do not have basic string lengths of
long periodic nucleotide strings, which are equal to the basic
periodic lengths. The basic periodic length $d_b \approx 135$ was
also found in the correlation analysis of the human
genomes\cite{RRP}.

Although more and more biological functions of the junk
DNA in cells are found, the mystery of transposable elements in
the whole genomes remains unraveled. The purpose of this work is to
depict the genome structure in the bacterial and
archaeal complete genomes and explain the genome dynamics in
terms of nucleotide string transfer. The proposed periodic
correlation structures with approximately the same increasing periods may have
fundamental importance for the biological functions of the junk
DNA.

\textbf{Acknowledgments} We would like to thank the National
Science Foundation for partial support through the Grant No.
11172310 and the IMECH/SCCAS SHENTENG 1800/7000 research computing facility for
assisting in the computation.

\newpage
{\footnotesize Table I. Basic strings and their lengths of long
periodic nucleotide strings ($ \geq 20$ bases) in bacterial and
archaeal complete genomes. $N_p$ is the number of the periods.
$N_s$ is the total number of the basic strings.

\begin{tabular}{llcl}
\hline
No & Genome(length) & $N_p/N_s$  & Periods(basic strings)\\
\hline
1 & ${mgen}(580074)$      & 1/4  & 3($tag$, $tgt$, $ct^2$, $cta$)\\
\hline
2 & ${mjan}(1664970)$     & 3/4  & 1($g$), 67($at^2 \cdots t^2c$), 68($at^2 \cdots atg$,$gt^2 \cdots a^2c$)\\
\hline

3 & ${hpyl}(1667867)$  & 8/14  & 6($cta^2gt$), 7($tgat^2ag$),
8($t^2gct^2ga$, $t^2atgtat$, $tca^2gca^2$),\\
& & & 12($ctc \cdots ctc$,$a^2t \cdots t^3$), 16($t^2a \cdots
tat$,
$a^2t \cdots ata$),\\
& & & 21($t^3 \cdots t^2c$,$tgt \cdots t^2c$,$tca \cdots tca$), 24($t^2c \cdots t^2c$),
390($t^3 \cdots a^2t$)\\
\hline

4& ${hpyl99}(1643831)$  & 10/19   & 1($a$),7($t^2agtga$),
8($ca^2c^2at^2$,$t^2atgtat$,$cat^2ca^2t$), 9($t^4gatga$,\\
& && $t^4ca^2tc$), 10($a^2gat^2a^3c$), 12($gt^2 \cdots
tgt$, $c^3 \cdots ta^2$,$tct \cdots atc$),\\
& &&  15($a^3 \cdots t^3$,
$t^2c \cdots gat$,$ca^2 \cdots ga^2$, $a^3 \cdots t^2c$),16($a^3 \cdots t^2c$),\\
& &&  21($atc \cdots atc$,$tgt \cdots t^2c$),
228($tcg \cdots ct^2$) \\
\hline

5 &${bbur}(910724)$   &2/3    & 60($tg^2 \cdots t^2g$), 162($gt^2 \cdots ct^2$, $gag \cdots ctg$)\\
\hline
6 &${rpxx}(1111523)$   &1/1    & 84($at^2 \cdots a^2t$) \\
\hline
7 &${hinf}(1830138)$   &6/12     &
3($at^2$),4($gtct$,$t^2ga$,$t^2g^2$,$tgac$,$a^2c^2$,$t^2gc$,$ca^2t$),
5($t^2atc$), \\
& &&9($cgc^2t^2gt^2$), 12($ct^2 \cdots gag$), 15($a^3 \cdots ag^2$) \\
\hline
9 &${mpneu}(816394)$   &1/3     & 12($t^2a \cdots cgc$,$tcg \cdots agt$, $ct^2 \cdots gca$)\\
\hline
10 &${mthe}(1751377)$  &3/3      & 5($cagtc$), 9($ct^2ct^2gta$), 44($tat \cdots ctg$)\\
\hline
14 & ${mtub}(4411529)$ &19/34       &9($t^2gtg^3c^2$,
$a^2cg^2cg^2c$, $g^2cg^2cac^2$), 15($tgc \cdots cac$), 30($gc^2
\cdots gtc$,\\
& & & $gt^2 \cdots cag$),
36($acg \cdots acg$, $ga^2 \cdots gca$), 51($tga \cdots gct$), \\
& && 53($ctc \cdots ctg$, $ctg \cdots gct$, $cat \cdots a^2c$),
54($agt \cdots gc^2$, $g^2c \cdots aca$), \\
& && 56($gc^2 \cdots tgc$, $gcg \cdots cga$, $c^2g \cdots ac^2$), 57($gtg \cdots tg^2$, $cta \cdots gct$, \\
& && $gca \cdots gct$, $gc^2
\cdots aca$), 58($gc^2 \cdots cga$), 59($agt \cdots cta$, $gcg \cdots cta$, \\
& && $cgc \cdots ca^2$), 63($gag \cdots agt$), 69($g^2c \cdots tc^2$),     75($agc \cdots gtc$,  \\
& && $c^2g \cdots c^2t$), 77($gat \cdots gct$), 78($g^2t \cdots gct$), 79($g^2t \cdots c^2a$), \\
& & & 111($gag \cdots c^2a$), 615($cg^2 \cdots cg^2$)\\
\hline
18 & ${ecoli}(4639221)$   &2/3      & 8($atga^3tg$, $gcactatg$), 113($cgc \cdots t^2a$)\\
\hline

19 & ${synecho}(3573470)$  &7/9     & 17($tat \cdots tgc$),
18($gag \cdots c^3$), 30($gag \cdots c^3$,$gag \cdots c^2t$), \\
& && 42($tca \cdots a^2t$), 78($gag \cdots c^3$), 318($gat \cdots at^2$), 888($cac \cdots cat$,\\
& & &  $a^2c \cdots cgt$) \\
\hline
22 & ${tpal}(1138011)$    &2/2    & 24($ctc \cdots t^2c$), 93($gct \cdots ga^2$)\\
\hline
\end{tabular}
}

\newpage
{\footnotesize Table II. Basic periodic lengths of periodic
correlation structures and basic string lengths of long periodic
nucleotide strings ($ \geq 20$ bases) in bacterial and archaeal
complete genomes
 denoted by $d_b$ and $p$, respectively.

\begin{tabular}{llcc}
\hline
No. & Genome(Acc. No.)      & $d_b$ & $p$\\
\hline
1 & ${mgen}(L43967)$       &3 & 3\\
\hline
2 & ${mjan}(L77117)$      &67 & 1, 67, 68\\
\hline
3 & ${hpyl}(AE000511)$      & 7,8, $\approx$ 114 & 6, 7, 8, 12, 16, 21, 24, 390\\
\hline
4 & ${hpyl99}(AE001439)$    & 8,15,21 & 1, 7, 8, 9, 10, 12, 15, 16, 21, 228 \\
\hline
5 & ${bbur}(AE000783)$      & 162 & 60, 162\\
\hline
6 & ${rpxx}(AJ235269)$      & 84 & 84 \\
\hline
7 & ${hinf}(L42023)$       & 4 & 3, 4, 5, 9, 12, 15 \\
\hline
8 & ${pNGR234}(U00090)$     & (6) & -\\
\hline
9 & ${mpneu}(U00089)$       & 12 & 12\\
\hline
10 & ${mthe}(AE000666)$      & 67 & 5, 9, 44\\
\hline
11 & ${aquae}(AE000657)$      & (67) & -\\
\hline
12 & ${pyro}(BA000001)$       & 67 & -\\
\hline
13 & ${aful}(AE000782)$       & 67 & -\\
\hline
14 & ${mtub}(AL123456)$       & 9,15,(57) &9, 15, 30, 36, 51, 53, 54, 56, 57, 58 \\
                            &        &           & 59, 63, 69, 75, 77, 78, 79, 111, 615\\
\hline
15 & ${pabyssi}(AL096836)$      & 67 & -\\
\hline
16 & ${tmar}(AE000512)$        & 67 & -\\
\hline
17 & ${cpneu}(AE001363)$        &(330) & -\\
\hline
18 & ${ecoli}(U00096)$        &100,113 & 8,113\\
\hline
19 & ${synecho}(BA000022)$      &6,888, $\approx$ 296& 17, 18, 30, 42, 78, 318, 888 \\
\hline
20 & ${ctra}(AE001273)$      &(108),(150) & -\\
\hline
21 & ${aero}(BA000002)$       & 65 & -\\
\hline
22 & ${tpal}(AE000520)$       & 24 & 24, 93\\
\hline
23 & ${bsub}(AL009126)$       & $\approx 5000$ & -\\
\hline
\end{tabular}
}

\newpage
\textbf{Figure caption}

Fig.~1. Plots of correlation intensity $\Xi(d)$ versus correlation
distance $d$ for (a) periodic and (b) random sequences.

Fig.~2. A plot of correlation intensity $\Xi(d)$ versus
correlation distance $d$ for the Methanococcus jannaschii DSM 2661 ({\it
mjan}) genome.

Fig.~3. A plot of correlation intensity $\Xi(d)$ versus
correlation distance $d$ for the Helicobacter pylori J99
 ({\it hpyl99}) genome.

Fig.~4. A plot of correlation intensity $\Xi(d)$ versus
correlation distance $d$ for the Bacillus subtilis subsp. subtilis str. 168 ({\it bsub})
genome.

Fig.~5. Plots of correlation intensity $\Xi(d)$ versus correlation
distance $d$ for (a) the Helicobacter pylori 26695 ({\it hpyl})
and (b) the Synechocystis sp. PCC6803 ({\it synecho}) genomes.

Fig.~6. A plot of correlation intensity $\Xi(d)$ versus
correlation distance $d$ for the Aquifex aeolicus VF5 ({\it aquae})
genome.
\end{document}